\documentclass[5p,twocolumn,number,sort&compress]{elsarticle}
\usepackage{dcolumn}
\usepackage{amsmath,amssymb,amsthm}
\usepackage{mathtools}
\usepackage{graphicx}
\usepackage{xcolor}
\usepackage{ascmac}
\usepackage[breaklinks=true,linktocpage=true]{hyperref} 
\usepackage{mathrsfs}
\newcommand{\ig}{\includegraphics}

\newcommand{\vevs}[1]{\langle #1 \rangle}

\newcommand{\link}{\,{\rm Link}\,}

\newcommand{\er}[1]{Eq.~\eqref{#1}}

\newcommand{\bb}{\mathbb}

\newcommand{\hph}{\hphantom}
\renewcommand{\b}{\bar}

\newcommand{\bs}{\boldsymbol}

\newcommand{\fr}{\frac}

\newcommand{\der}{\partial}

\newcommand{\wed}{\wedge}

\begin{document}
\allowdisplaybreaks[4]

\title{Topological axion electrodynamics and 4-group symmetry}

\author[1,2,3]{Yoshimasa Hidaka}
\ead{hidaka@post.kek.jp}
\author[4]{Muneto Nitta}
\ead{nitta@phys-h.keio.ac.jp}
\author[1,4]{Ryo Yokokura}
\ead{ryokokur@post.kek.jp}
\address[1]{KEK Theory Center, Tsukuba 305-0801, Japan}
\address[2]{Graduate University for Advanced Studies (Sokendai), 
Tsukuba 305-0801, Japan}
\address[3]{RIKEN iTHEMS, RIKEN, Wako 351-0198, Japan}
\address[4]{Department of Physics \& Research and Education Center for
Natural Sciences, Keio University, 
\par
Hiyoshi 4-1-1, Yokohama, Kanagawa
223-8521, Japan}
\tnotetext[t0]{KEK-TH-2331, J-PARC-TH-0243}

\date{\today}

 \setcounter{footnote}{0}%
\renewcommand{\thefootnote}{$*$\arabic{footnote}}%
\begin{abstract}
We study higher-form symmetries and a higher group in 
the low energy limit of a $(3+1)$-dimensional 
axion electrodynamics with a massive axion and a massive photon.
A topological field theory 
describing 
topological excitations with the axion-photon coupling, 
which we call a topological axion electrodynamics,
is obtained in the low energy limit.
Higher-form symmetries of the topological axion electrodynamics 
are specified by equations of motion and Bianchi identities.
We find that there are induced anyons on the intersections
of symmetry generators.
By a link of worldlines of the anyons, 
we show that the worldvolume of an 
axionic domain wall is topologically ordered.
We further specify the underlying mathematical structure elegantly describing 
all salient features of the theory to be a 4-group.
\end{abstract}

\maketitle

\section{Introduction}
Axions are hypothetical particles 
originally proposed as 
 a solution to the strong CP problem~\cite{Peccei:1977hh,Weinberg:1977ma,Wilczek:1977pj,Dine:1981rt,Zhitnitsky:1980tq,Kim:1979if,Shifman:1979if} in particle physics, 
 but now play crucial roles in  several contexts in modern physics;
they are not only a dark matter candidate in cosmology~\cite{Preskill:1982cy,Abbott:1982af,Dine:1982ah,Stecker:1982ws,Masso:1995tw}
(see, e.g., Refs.~\cite{Kim:1986ax,Dine:2000cj,Peccei:2006as,Kawasaki:2013ae}
as a review) 
but also appear in string theories~\cite{Witten:1985xb,Townsend:1993wy,Izquierdo:1993st,Harvey:2000yg,Svrcek:2006yi,Arvanitaki:2009fg} 
and even in topological 
insulators~\cite{Wilczek:1987mv,Qi:2008ew,Essin:2008rq} 
and topological superconductors~\cite{Qi:2012cs,Stone:2016pof,Stalhammar:2021tcq} 
in condensed matter physics. 
(see, e.g., Refs.~\cite{Hasan:2010xy,Sekine:2020ixs} as a review).
One of the salient features of the axions 
is 
a topological coupling to 
the photon due to a chiral anomaly.
In particular, the simplest system, the axion electrodynamics~\cite{Wilczek:1987mv}, 
has been studied extensively to capture
magneto-electric responses due to the topological 
coupling~\cite{Fischler:1983sc,Sikivie:1984yz,
Kaplan:1987kh,Manohar:1988gv,Wilczek:1987mv,Kogan:1992bq,Kogan:1993yw,
Qi:2008ew,Essin:2008rq,
Ferrer:2015iop,Yamamoto:2015maz,Ferrer:2016toh}.
When both the axion and photon have mass gaps, the axion electrodynamics is applicable to topological 
superconductors~\cite{Qi:2012cs,Stone:2016pof,Stalhammar:2021tcq} 
and admits topological solitons such as 
axionic domain walls~\cite{Sikivie:1982qv,Vilenkin:1982ks} 
and quantized magnetic 
vortex strings~\cite{Abrikosov:1956sx,Nielsen:1973cs}.

Topological solitons are necessary ingredients to determine the dynamics and phases of systems.
In the absence of the axion coupling, 
quantized magnetic vortices can give 
fractional Aharonov-Bohm (AB) effects.
Gapped phases exhibiting these effects 
in the low energy limit
are called topologically ordered 
phases~\cite{Wen:1989iv,Wen:1990zza,Wen:1991rp,Hansson:2004wca},
which are realized by anyons in fractional quantum Hall (FQH) states
and s-wave superconductors in $(2+1)$ and $(3+1)$ dimensions, respectively. 
Once we take into account the axion coupling, electromagnetic 
properties become richer: for instance, 
electric charges are 
induced by penetrating magnetic fluxes to the axionic domain wall
~\cite{Sikivie:1984yz,Wilczek:1987mv,Qi:2008ew,Teo:2010zb}.
Effects on the domain wall can be described by an Abelian Chern-Simons (CS) term on the domain wall, 
induced by the axion-photon 
coupling~\cite{Lee:1986mm,Komargodski:2018odf}.
Since FQH states can be described by this CS 
term in the low energy limit~\cite{Wen:1989iv,Wen:1990zza,Wen:1991rp},
one can expect that the axionic domain wall is topologically ordered.
However, topological objects giving the AB phases, 
which we will call topological order parameters in this Letter, 
have not been identified to the best of our knowledge.

On the other hand, recently physics of topological solitons and extended objects has been 
extensively studied in the language of global higher-form symmetries;
symmetries under actions on $p$-dimensional extended objects, called $p$-form 
symmetries~\cite{Banks:2010zn,Kapustin:2014gua,Gaiotto:2014kfa,Batista:2004sc, Pantev:2005zs,Pantev:2005wj,Pantev:2005rh,Nussinov:2006iva, Nussinov:2008aa, Nussinov:2009zz, Nussinov:2011mz,Distler:2010zg}, 
were found as natural extensions of ordinary symmetries
acting on local 0-dimensional operators.
Higher-form symmetries give us 
new understandings for the classification of phases of matter 
and physical effects discussed 
by extended objects, e.g.,
topologically ordered phases as spontaneous symmetry breaking 
of 1-form symmetries associated to the AB effect~\cite{Nussinov:2009zz,Gaiotto:2014kfa,Hidaka:2019jtv},
the magneto-electric responses 
for the axion electrodynamics in gapless phase~\cite{Hidaka:2020iaz,Hidaka:2020izy}
as correlations of 
0- and 1-form symmetries.
One of the most elegant notions of symmetries  
in modern quantum field theory 
can be formulated as so-called higher groups;
 an $n$-group symmetry denotes 
a set of 0-, ..., $(n-1)$-form symmetries 
with nontrivial correlations between them. 
2- and 3-groups have been 
extensively studied (see e.g., 
Refs.~\cite{Sharpe:2015mja,Tachikawa:2017gyf,Cordova:2018cvg,1707352,Bhardwaj:2016clt, Kapustin:2017jrc,deAlmeida:2017dhy,Benini:2018reh, Delcamp:2018wlb, Wen:2018zux, Delcamp:2019fdp, Thorngren:2020aph,Cordova:2020tij,Hsin:2019fhf,Gukov:2020btk,Iqbal:2020lrt,Brauner:2020rtz,DeWolfe:2020uzb,Brennan:2020ehu,Heidenreich:2020pkc}), 
and in particular, 
 a 3-group structure has been found in gapless axion electrodynamics~\cite{Hidaka:2020iaz,Hidaka:2020izy}.

In this Letter, we investigate higher-form global symmetries and a higher group
in the low energy limit of 
the $(3+1)$-dimensional axion electrodynamics in the gapped phase, 
to show a topological order on the axionic domain wall.
The higher-form symmetries can be specified
by employing a dual topological quantum field theory
for the massive 
photon~\cite{Cremmer:1973mg,Davis:1988rw,Horowitz:1989ng,Blau:1989bq,Allen:1990gb}
and massive axion~\cite{Aurilia:1977jz,Aurilia:1980jz,Dvali:2005an,Kaloper:2008fb,Hidaka:2019mfm}, which we will call a {\it topological axion electrodynamics}.
We find 0-, 1-, 2-, and 3-form symmetries 
and show the topological order on the axionic domain wall 
in terms of the higher-form symmetries,
by identifying the topological order parameters as intersections of 
0- and 1-form symmetry generators on which anyons are induced.
On the axionic domain wall, 
the intersections can have a fractional phase
determined by the groups of the higher-form symmetries. 
The fractional AB effect for 
the intersected symmetry generators 
implies nontrivial correlations between higher-form symmetries. 
We further find that the symmetries organize a 4-group symmetry.
While a 4-group symmetry in a simpler case has been studied 
in Ref.~\cite{Tanizaki:2019rbk},
our 4-group is the first example where all the 0-, 1-, 2-, and 3-form 
symmetries are nontrivial.

\section{Topological axion electrodynamics}
Here, we give an action of the low energy limit of
 the axion electrodynamics 
with a massive axion and a massive photon.
Using dual transformations,
we derive a topological field theory describing this limit of the system.
We begin with the effective action,
\begin{equation}
\begin{split}
S
&= 
- \int_{M_4} \Big(
\fr{v^2}{2}|d\phi|^2 + \fr{1}{2e^2} |da|^2 + \fr{v'^2}{2}|d \chi - q a |^2
\\
&\quad
\hph{- \int_{M_4} \Big(}
+ V(\phi)\star 1   
-\fr{N}{8\pi^2} \phi da \wed da \Big), 
\end{split}
\label{210424.1623}
\end{equation}
where we use notations of differential forms: 
$|\omega_p|^2  = \omega_p \wed \star \omega_p$
for a $p$-form $\omega_p$, and $\star$ is the Hodge star operator.
$M_4$ denotes the $(3+1)$-dimensional spacetime spin 
manifold~\cite{Alvarez:1999uq,Olive:2000yy}.
The quantities $v,v'$ 
are mass dimension 1 parameters, $e$ is a coupling constant,
and $N$ is an integer.
The axion $\phi$ is given by a $2\pi$ periodic pseudo-scalar field,
$ \phi ({\cal P})+ 2\pi \sim \phi ({\cal P})$, 
for a point ${\cal P}$ in the spacetime.
By this periodicity, the axion can have a winding number along 
a loop ${\cal C}$:
$ \int_{\cal C} d\phi \in 2\pi \bb{Z}$.
In addition, we have introduced a potential term 
$V(\phi)$ for the axion, 
which has 
a global symmetry under the shift 
$ V\left(\phi + {2\pi}/{k}\right) = V(\phi)$.
The potential has $k$ of degenerated minima,
$V({2\pi n}/{k}) = V'({2\pi n}/{k}) = 0$ and $V''({2\pi n}/{k}) >0$
for $n \in \bb{Z}$ mod $k$. 
The photon is described by a $U(1)$ 1-form gauge field $a$
with a gauge transformation,
$ a \to a + d\lambda$.
Here, $\lambda$ is a $U(1)$ 0-form gauge parameter normalized as $ \int_{\cal C} d\lambda \in 2\pi \bb{Z}$.
The photon is subject to the Dirac quantization
on a closed 2-dimensional subspace ${\cal S}$ (e.g., a sphere $S^2$),
$ \int_{\cal S} da \in 2\pi \bb{Z}$.
The $2\pi $ periodic scalar field $\chi$ is introduced as  
a phase component of a charge $q (\in \bb{Z})$ Higgs field. 
The field $\chi $ is shifted as $ \chi \to \chi + q \lambda$ under 
the gauge transformation
$ a \to a + d\lambda$.

To investigate topological properties of the system, 
we dualize the theory to a topological theory
after neglecting the kinetic terms of the axion and photon
at low energy.
We dualize 
$|d\chi - qa|^2$
and 
$V(\phi)$
to topological term given by 
2- and 3-form gauge fields,
respectively~\cite{Cremmer:1973mg,
Davis:1988rw,Horowitz:1989ng,Blau:1989bq,Allen:1990gb,Aurilia:1977jz,
Aurilia:1980jz,Dvali:2005an,Kaloper:2008fb,
Hidaka:2019mfm}.
The dual topological action has 
the simple form, 
\begin{equation}
    S_{\rm TAE}
= 
\int_{M_4} \left( \fr{k}{2\pi} c \wed  d\phi 
+ \fr{q}{2\pi} b \wed da
+
\fr{N}{8\pi^2} \phi da \wed da\right) .
\label{210514.1613}
\end{equation}
Here, $b$ and $c$ are $U(1)$ 2- and 3-form gauge fields,
whose gauge transformation laws are
given by 1- and 2-form gauge parameters $\lambda_1$ and $\lambda_2$ as
$ b \to b +d\lambda_1$
and $c \to c+ d\lambda_2$,
respectively.
They are normalized by the flux quantization conditions as
$ \int_{\cal V} db , \, \int_{\Omega} dc, \int_{\cal S} d\lambda_1, 
\int_{\cal V} d\lambda_2\in 2\pi \bb{Z}$
where ${\cal V}$ and 
$\Omega$ are closed 3- and 4-dimensional subspaces, respectively.
We will call the dual theory 
the ``topological axion electrodynamics,''
since it can be described by only topological terms.

\section{Higher-form global symmetries in topological axion electrodynamics}
We show higher-form global symmetries in this system 
found by the equations of motion 
for the dynamical fields $\phi$, $a$, $b$, and $c$:
\begin{equation}
\begin{split}
& \frac{k}{2\pi} d c + \fr{N}{8\pi^2} da \wed da =0,
\;\;\,
\frac{q}{2\pi} db +\fr{N}{4\pi^2} d\phi \wed da =0, 
\\
&\frac{q}{2\pi} da =0, 
\quad
\frac{k}{2\pi} d\phi =0, 
\end{split}
\label{210624.2012}
\end{equation}
respectively.
The corresponding symmetry generators
with groups parametrizing them are
\begin{equation}
\begin{split}
  U_0 (e^{2\pi i n_0/m},{\cal V})
&= e^{ -i \fr{n_0}{m} \int_{\cal V} (kc + \fr{N}{4\pi} a \wed da)},
\\
 U_1(e^{2\pi i n_1/p}, {\cal S})
& = e^{ -i \fr{n_1}{p} \int_{\cal S} (q b + \fr{N}{2\pi}\phi da)},
\\
 U_2 (e^{2\pi i n_2/q},{\cal C})
&= e^{- in_2 \int_{\cal C} a},
\\
 U_3 (e^{2\pi i n_3/k},{\cal (P,P')})
&= e^{ -in_3 (\phi ({\cal P})-\phi({\cal P'}) )}.
\end{split}
\label{210509.1817}
\end{equation}
Here, we have introduced 
$n_0 ,..., n_3 \in \bb{Z}$,
$ p \coloneqq \gcd (N, q)$, and $m \coloneqq \gcd(N,k)$,
where ``$\gcd$'' represents the greatest common divisor.
Hereafter, we assume that the subspaces ${\cal V, S}$, and ${\cal C}$ 
do not have any self-intersection for simplicity.
All symmetry generators have the standard form of generators $U_{n}=e^{i\theta_{n}Q_{n}}$ with $Q_{n}= \int j_{n}$, where the $(3-n)$-form current $j_{n}$ is closed under the equations of motion~\eqref{210509.1817}, i.e., $dj_{n}=0$.
Note that the phase $\theta_{n}$ is 
constrained by 
the large gauge invariance of the dynamical 
fields~\cite{Henneaux:1986tt,Dijkgraaf:1989pz,Witten:2003ya,Armoni:2018bga,Hidaka:2020iaz,Hidaka:2020izy}. 
Therefore, there are
the electric
$\bb{Z}_m$ 0-form, $\bb{Z}_p$ 1-form, $\bb{Z}_q$ 2-form, 
and $\bb{Z}_k$ 3-form global symmetries.
Their physical interpretations
are as follows:
$ U_0 (e^{2\pi i n_0/m},{\cal V})$ 
is a worldvolume of the axionic domain wall:
the equation of motion of $c$ in the presence of $U_0 $
is $d\phi = ({2\pi n_0}/m)\delta_1 ({\cal V})$.
Here, $\delta_{4-n} (\Sigma_n)$ is a delta function $(4-n)$-form 
satisfying 
$\int_{\Sigma_n} \omega_n = \int_{M_4} \omega_{n} \wed \delta_{4-n}(\Sigma_n)$ for a $n$-dimensional subspace $\Sigma_n$ 
\cite{Horowitz:1989km,Oda:1989tq,Chen:2015gma,Hidaka:2019jtv,Yamamoto:2020vlk}. 
Similarly, 
$ U_1(e^{2\pi i n_1/p}, {\cal S})$ 
represents a worldsheet of a quantized magnetic flux
or an impulse of a quantized electric field,
because the equation of motion of $b$ in the presence of $U_1$ 
is $da = (2\pi n_1 /p) \delta_2 ({\cal S})$.
$U_2$ and $U_3$ are a Wilson loop and a 2-point local 
operator of the axion, respectively.
All of them are topological, and the 
vacuum expectation values (VEVs) are trivial,
$\vevs{U_0} = \cdots = \vevs{U_3} =1$
thanks to the equations of motion 
in \er{210624.2012}.
Here, the symbol `$\vevs{\cdots}$' means the VEV.
We remark that the 2- and 3-form global symmetries 
are emergent symmetries in the low energy limit,
where the magnetic vortices and axionic domain walls
can be treated as objects with infinitely large tensions. 
We summarize corresponding charged objects,
which are not necessary in the following discussion
in \ref{charge}. 

In addition to these electric symmetries, there are 
four magnetic $U(1)$ $(-1)$-, 0-, 1-, 2-form symmetries,
which are given as the form
$U_n^M(e^{i\theta^M_{n}},\Sigma_{3-n})=e^{i\theta^M_n\int_{\Sigma_n} j^M_n}$ with 
closed currents $j^M_{-1}=dc$, $j^M_0=db$, $j^M_1=da$, $j^M_2=d\phi$.
The conservation laws are given by the Bianchi identities.
In this Letter, we mainly focus on the electric symmetries.

\section{Topological order in bulk}
Before discussing the topological order on the axionic domain wall, we show the topological order in the bulk for $p = \gcd(N,q) \neq 1$
with a fractional AB phase given by $\bb{Z}_p$.
Topological order parameters can be identified as 
the symmetry generators $U_1$ and $U_2$, 
since they are topological and have nontrivial fractional phases.
First, we have seen that 
they are topological, developing nonzero VEVs
$\vevs{ U_2 (e^{2\pi i n_2/q}, {\cal C})} =
\vevs{ U_1 (e^{2\pi i n_1/p}, {\cal S})} = 1$.
Second, they have a fractional AB phase
characterized by $\bb{Z}_p$:
\begin{equation}
\vevs{U_1 (e^{\fr{2\pi i n_1}{p}}, {\cal S}) 
U_2 (e^{\fr{2\pi i n_2}{q}}, {\cal C})}
=
e^{i\theta_{12}}.
\label{210516.1648}
\end{equation}
Here, we have defined 
$\theta_{12} = -(2\pi {n_1 n_2 }/{p}) \link ({\cal S,C}) $
with
the linking number between 
$n$- and $(3-n)$-dimensional subspaces 
$\Sigma_{n}$ and $  \Sigma'_{3-n}$ 
as  
$\link (\Sigma_{n} , \Sigma'_{3-n}) 
= \int_{\Omega_{\Sigma_n}}\delta_{n+1} (\Sigma'_{3-n})$
by using $(n+1)$-dimensional subspace $\Omega_{\Sigma_n}$ satisfying
 $\der \Omega_{\Sigma_n} = \Sigma_n$ (see \ref{corr}).
By the linking phase, 
the topological axion electrodynamics is topologically ordered in the bulk.
We remark that this fractional AB phase is different from a $\bb{Z}_q$ AB phase of an Abelian Higgs model without an axion
\cite{Hansson:2004wca,Gaiotto:2014kfa}, 
due to the presence of the axion-photon coupling 
deforming the 1-form symmetry.

\section{Topological order on axionic domain walls}
Now, we show that the axionic domain walls are also topologically
ordered with a nontrivial fractional linking phase 
in a different manner from the bulk.
The topological order parameter
can be identified 
as an intersection of the 0- and 1-form symmetry generators. 
The fractional linking phase is given by 
the group $\bb{Z}_{mp^2/\gcd(N,mp^2)}$, 
different from the one in the bulk.

A rough description is as follows. 
The worldvolume of the axionic domain wall 
can be understood as a FQH state, since
$U_0$ in \er{210509.1817} has a level $N/m \in \bb{Z}$ 
CS term, 
$\fr{N}{4\pi m } \int_{\cal V} a \wed da$.
By intersecting $U_1$ to $U_0$, we can have an anyon, whose worldline 
is a topological order parameter of FQH states.
Linked anyons on the domain wall 
can be obtained by two $U_1$'s intersected in the bulk.

\begin{figure}[t]
 \centering
\ig[height=7em]{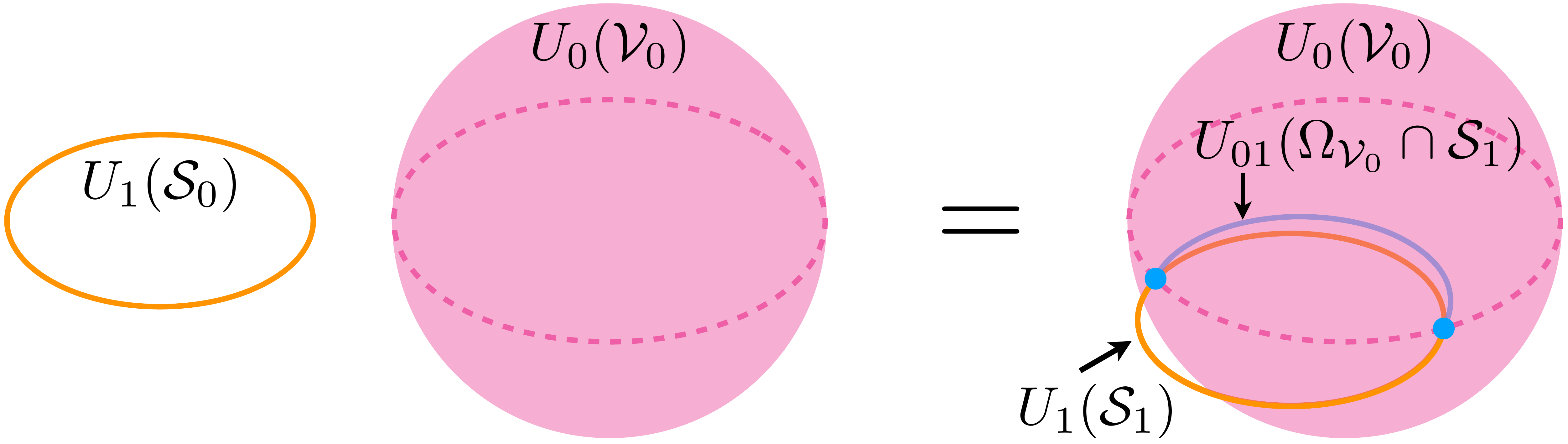}
\caption{\label{U01}Intersection of 
temporally and spatially extended
symmetry generators $U_0 $ (pink sphere) 
and $U_1 $ (orange line).
This figure is a time slice of the configuration.
The blue line in the right panel is a time slice of an induced static surface $U_{01}$.
The blue dots denote the boundary of $\Omega_{{\cal V}_0}\cap {\cal S}_1$ on the time slice,
which corresponds to an anyon on the domain wall.
}
\end{figure}

A more precise proof can be given as follows.
First, we intersect $U_0 $ to $U_1$ to create a worldline of an anyon on the domain wall.
We begin with the 
correlation function
$
\vevs{U_0 (e^{2\pi i n_0/m}, {\cal V}_0) 
U_1 (e^{2\pi i n_1/p}, {\cal S}_0)}$ 
assuming
${\cal V}_0 \cap {\cal S}_0 = \emptyset$
(the left panel of Fig.~\ref{U01}).
We deform 
${\cal S}_0 $ to ${\cal S}_1$
that is intersected to ${\cal V}_0$
(the right panel of Fig.~\ref{U01}).
Here, we choose ${\cal S}_1$ so that 
$ {\cal S}_1\cap {\cal V}_0$ is a 1-dimensional 
closed subspace,
and ${\cal S}_0 \cap {\cal S}_1 = \emptyset$.
We interpolate between ${\cal S}_0$ and ${\cal S}_1$ using 
a 3-dimensional subspace ${\cal V}_{01}$
satisfying $\der {\cal V}_{01} = {\cal S}_0 \cup \b{\cal S}_1$.
We assume that ${\cal V}_{01}$ does not intersect with 
any singularity such as an 't Hooft line.
Using $U_1 (e^{2\pi i n_1/p}, {\cal S}_0)=U_1 (e^{2\pi i n_1/p},  \der {\cal V}_{01} )U_1 (e^{2\pi i n_1/p}, {\cal S}_1)$,
and absorbing $U_1 (e^{2\pi i n_1/p},  \der {\cal V}_{01} )$ 
into the action by a redefinition 
$a + ({2\pi n_1}/{p}) \delta_1 ({\cal V}_{01}) \to a$,
the correlation function can be written as
\begin{equation}
\begin{split}
& \vevs{U_0 (e^{2\pi i n_0/m}, {\cal V}_0) 
U_1 (e^{2\pi i n_1/p}, {\cal S}_0)}
\\
&
= 
 \vevs{
 U_{01} (e^{\fr{2\pi i N n_0n_1}{mp}}, \Omega_{{\cal V}_0} \cap {\cal S}_1)
\\
&
\qquad
\times 
U_0 (e^{2\pi i n_0/m}, {\cal V}_0) 
U_1 (e^{2\pi i n_1/p}, {\cal S}_1)}.
\end{split}
\label{210628.2349}
\end{equation}
Here, we have introduced 
an object on 
a 2-dimensional subspace, $\Omega_{{\cal V}_0} \cap {\cal S}_1$, with a boundary,
\begin{equation}
  U_{01} (e^{\fr{2\pi i N n_0n_1}{mp}}, \Omega_{{\cal V}_0} \cap {\cal S}_1)
= 
e^{\fr{ 2\pi i N n_0n_1 }{mp} \int_{\Omega_{{\cal V}_0}} \fr{da}{2\pi} \wed 
\delta_2 ({\cal S}_{1})},
\label{eq:U01}
\end{equation}
where 
$\Omega_{{\cal V}_0}$ is a 4-dimensional subspace 
whose boundary is ${\cal V}_0$.
It is straightforward to show that $U_{01}$
is topological under the deformation of ${\cal S}_1$ or ${\cal V}_0$ 
by the redefinition of $a$ or $\phi$, respectively.
Therefore, we find that, when $U_0$ and $U_1$ are intersected, there should be an additional magnetic 1-form symmetry generator with the boundary $U_{01}$.
The boundary object expresses a worldline of an anyon on the domain wall.
This anyon cannot solely exist in the bulk if $N/ mp \not\in \bb{Z}$; 
it is always trapped on the domain wall.
From the phase factor in Eq.~\eqref{eq:U01}, the anyon line has a fractional electric charge ${ N n_0n_1 }/(mp)$~\cite{Sikivie:1984yz}.
The appearance of anyons trapped on the domain wall is one of the main results of this Letter.
The anyons also have a fractional linking phase, which we will show in the following.

\begin{figure}[t]
 \centering
\ig[height=7em]{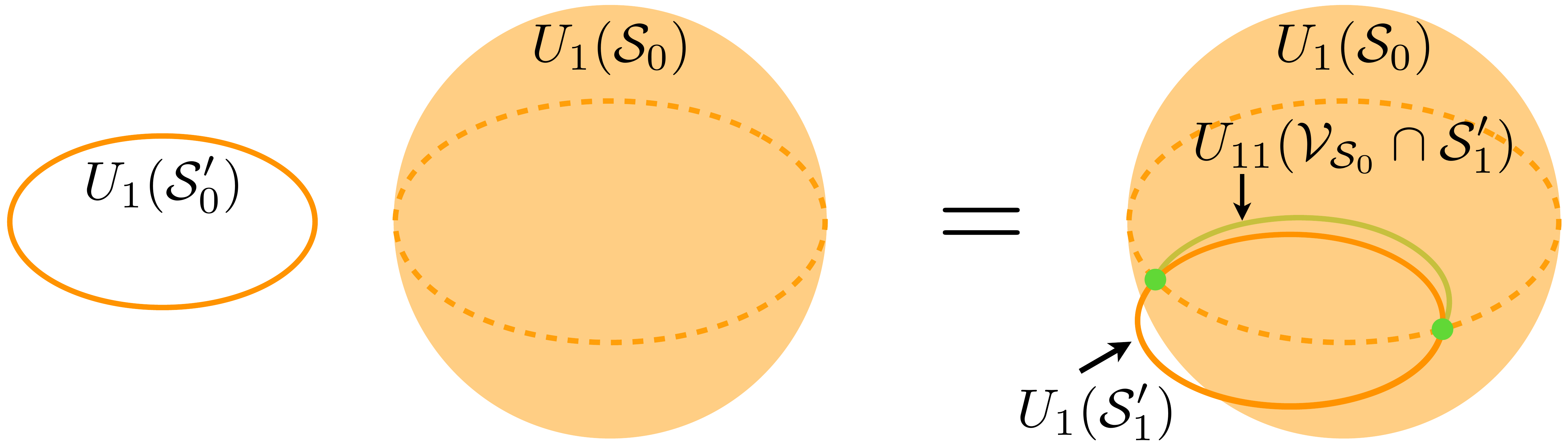}
\caption{\label{U11}
Intersection of symmetry generators $U_1$.
This figure is a time slice of the configuration:
$U_1({\cal S}_0)$ expressed by orange spheres
are extended to only spatial directions,
while 
$U_1({\cal S}_0')$ and $U_1({\cal S}_1')$ 
expressed by the orange lines are 
temporally and spatially extended.
The green line in the right panel is an induced instantaneous line object $U_{11}$.
The green dots denote boundaries of ${\cal V}_{{\cal S}_0} \cap {\cal S}'_1$
on the time slice.
}
\end{figure}

We next consider intersections of 1-form symmetry generators, which 
are necessary to derive the fractional linking phase of anyons.
We begin with the correlation function
(left panel of Fig.~\ref{U11}),
$\vevs{
U_1 (e^{2\pi i n_1/p}, {\cal S}_0)
U_1 (e^{2\pi i n'_1/p}, {\cal S}'_0)}$,
where 2-dimensional subspaces 
${\cal S}_0$ and ${\cal S}_0'$ 
satisfy ${\cal S}_0 \cap {\cal S}_0' = \emptyset$.
We deform ${\cal S}'_0$ to ${\cal S}'_1$ that intersects with 
${\cal S}_0$,
where the deformation is characterized by 
a 3-dimensional subspace ${\cal V}_{01}$ satisfying
$\der {\cal V}'_{01} = {\cal S}'_0 \cup \b{\cal S}_1'$.
By the same procedure as \er{210628.2349},
the correlation function becomes
\begin{equation}
\begin{split}
&
\vevs{
U_1 (e^{2\pi i n_1/p}, {\cal S}_0)
U_1 (e^{2\pi i n'_1/p}, {\cal S}'_0)}
\\
& =  
\vevs{
U_{11}( e^{ \fr{2 \pi i N n_1n_1' }{p^2} },
{\cal V}_{{\cal S}_0} \cap {\cal S}_{1}')
\\
&
\qquad \times
U_1 (e^{2\pi i n_1/p}, {\cal S}_0)
U_1 (e^{2\pi i n'_1/p}, {\cal S}'_1)
},
\end{split}
\label{eq:U1U1}
\end{equation}
where 
we have 
introduced
an object on a 1-dimensional subspace, ${\cal V}_{{\cal S}_0} \cap {\cal S}_{1}'$, with boundaries,
\begin{equation}
 U_{11}( e^{ \fr{2 \pi i N n_1n_1' }{p^2} },
{\cal V}_{{\cal S}_0} \cap {\cal S}_{1}') =
 e^{ \fr{2 \pi i N n_1n_1' }{p^2} 
\int_{{\cal V}_{{\cal S}_0}}
 \fr{d\phi}{2\pi} \wed \delta_2 ({\cal S}_{1}')},
\end{equation}
and 
${\cal V}_{{\cal S}_0}$ is a 3-dimensional subspace 
whose boundary is ${\cal S}_0$. 
Thus, we should add the magnetic 2-form symmetry generator with boundaries $U_{11}$ 
if we intersect two $U_1$'s.
Again, the intersected configuration is topological 
under deformations of ${\cal S}_0$ or ${\cal S}_1'$.
Using the interpretation of 
$U_1$ as external electric or magnetic fluxes,
the presence of $U_{11}$ can be physically understood as the fact that the $\bs{E} \cdot \bs{B}$ is a source of the axion,
since the boundary objects of $U_{11}$ can be
identified as local objects of the axion.

Finally, we consider the intersection of three symmetry 
generators to show the fractional linking phase of anyons on the domain wall. 
We begin with the following cubic but trivial correlation 
function,
\begin{equation}
\begin{split}
 \vevs{U_1 (e^{\fr{2\pi i n_1}{p}},{\cal S}_0)
 U_1 (e^{\fr{2\pi i n_1'}{p}},{\cal S}'_0)
 U_0 (e^{\fr{2\pi i n_0}{m}},{\cal V}_0)}=1,
\end{split}
\end{equation}
where we choose the subspaces such that 
${\cal S}_0 \cap {\cal S}'_0 = {\cal S}_0 \cap {\cal V}_0 = 
{\cal S}'_0 \cap {\cal V}_0 = \emptyset$.
We deform ${\cal S}_0$ to ${\cal S}_1$ using ${\cal V}_{01}$,
where ${\cal S}_1$ intersects with ${\cal V}_0$,
but ${\cal S}_1 \cap {\cal S}_0' = \emptyset$.
As shown in \er{210628.2349}, we have 
$U_{01}(e^{\fr{2 \pi i N n_0n_1}{mp}}, \Omega_{{\cal V}_0}
\cap {\cal S}_1)$
in the correlation function.
We then deform ${\cal S}_0'$ to ${\cal S}'_1$ interpolated by ${\cal V}'_{01}$, where
${\cal S}_1'$ intersects with ${\cal V}_0$, 
and also with ${\cal S}_1$ transversally.
We thus obtain
\begin{align}
&C_{011}(e^{2\pi i n_0/m},{\cal V}_0; 
e^{2\pi i n_1/p},{\cal S}_1; e^{2\pi i n'_1/p},{\cal S}_1')
\nonumber
\\
&
:=
\langle
U_{01}(e^{\fr{2 \pi i N n_0n_1}{mp}}, \Omega_{{\cal V}_0}
\cap {\cal S}_1)
U_{01}(e^{\fr{2 \pi i N n_0n'_1}{mp}}, \Omega_{{\cal V}_0}
\cap {\cal S}'_1)
\nonumber
\\
&
\quad
\times 
U_{11}(e^{ \fr{2 \pi i N n_1n_1'}{p^2} } , 
{\cal V}_{{\cal S}_1}\cap {\cal S}_{1}')
U_0 (e^{2\pi i n_0/m}, {\cal V}_0) 
\label{210630.0330}
\\
&
\quad
\times 
U_1 (e^{2\pi i n_1/p}, {\cal S}_1)
U_1 (e^{2\pi i n_1'/p}, {\cal S}'_1)
\rangle
= e^{i\theta_{011}}.
\nonumber
\end{align}
Here, 
$\theta_{011}
:=
-2\pi{N}n_0 n_1n_1'/(mp^2 )
\link ({\cal S}_1,{\cal S}'_1)|_{{\cal V}_0}
$,
and 
we have used 
the linking number of ${\cal S}_1$ and ${\cal S}'_1$
on ${\cal V}_0$
$
\link ({\cal S}_1,{\cal S}'_1)|_{{\cal V}_0}
:=\int_{{\cal V}_0}  
\delta_1({\cal V}_{{\cal S}_1}) \wed 
\delta_2 ({\cal S}_1') 
$,
and ${\cal V}_{{\cal S}_1}$ is a 3-dimensional subspace 
whose boundary is ${\cal S}_1$.
The above relation means that 
the boundary line objects of $U_{01}$
are topological order parameters with the 
fractional linking phase.
Thus, the axionic domain wall is topologically ordered,
and the fractional phase, 
$e^{i\theta_{011}} \in \bb{Z}_{mp^2/\gcd(N, mp^2)}$,
is different from that of the bulk.
The linking phase comes from the
fractionally quantized magnetic fluxes $(1/p)^2$ on
the level $N/m$ Abelian CS action.

\section{Global 4-group symmetry in topological axion electrodynamics}
The topological order on the axionic domain wall
implies a so-called higher-group structure; 
as shown below, 
there is a 4-group symmetry  
which is a set of 0-, 1-, 2-, and 3-form symmetries with nontrivial 
correlations between them, 
where the 0- and 1-form symmetry generators 
lead to a 2-form symmetry generator, 
and two 1-form symmetry generators lead to 
a 3-form symmetry generator.

As shown in Eq.~\eqref{210628.2349},
the correlation of the 0- and 1-form symmetry generators
induces a magnetic 1-form symmetry generator, whose boundary
is the intersection of the 0- and 1-form symmetry generators.
The intersection is a closed 1-dimensional object, which 
generate a 2-form symmetry. To see this, 
we again focus on \er{210630.0330},
and evaluate it as follows:
\begin{equation}
 \begin{split}
&C_{011}(e^{2\pi i n_0/m},{\cal V}_0; 
e^{2\pi i n_1/p},{\cal S}_1; e^{2\pi i n'_1/p},{\cal S}_1')
\\
&
=
 e^{i\theta_{011}}
\vevs{U_1 (e^{2\pi i n_1/p}, {\cal S}_1)
}.
 \end{split}
\label{210630.0345}  
\end{equation}
The 2-form symmetry and 
its symmetry group can be identified as follows.
We remark that 
$U_1$ is charged under the 2-form symmetry  
with the charge $ -n_1 q/p$,
$
\vevs{ U_2 (e^{\fr{2\pi i n_2}{q}}, {\cal C})
 U_1 (e^{\fr{2\pi i n_1}{p}}, {\cal S})}
=
e^{i\theta_{12}}
\vevs{ U_1 (e^{\fr{2\pi i n_1}{p}}, {\cal S})}
$,
as in \er{210516.1648}.
Thus, the intersection of $U_0$ and $U_1$ 
can be regarded as a 2-form symmetry generator.
By comparing $\theta_{011}$ to the 
charge $-n_1 q/p$,
we find that 
the intersection of $U_0$ and $U_1$
is parametrized by 
$\bb{Z}_{Q}$
with 
$Q = q\cdot mp / \gcd(N, mp)$ being product of 
$q$ and the
denominator of $N/ mp$.
Thus, the 2-form symmetry group is
identified as $\bb{Z}_{Q}$, which is transmuted 
from $\bb{Z}_q$.

Similarly, the intersection of two 1-form symmetry generators 
becomes the boundary of a magnetic 2-form symmetry~\eqref{eq:U1U1},
which behaves as a 3-form symmetry generator.
We again focus on the correlation function in \er{210630.0330},
and evaluate it as follows:
\begin{equation}
\begin{split}
&
C_{011}(e^{2\pi i n_0/m},{\cal V}_0;e^{2\pi i n_1/p},{\cal S}_1; e^{2\pi i n'_1/p},{\cal S}'_1)
\\
&
=
e^{i\theta_{011}
}
\vevs{ U_0 (e^{2\pi i n_0/m},{\cal V}_0)}
.
 \end{split}
\label{210606.1656}
\end{equation}
Since  
the correlation function
$
\langle
U_3 (e^{2\pi i n_3/k}, {\cal (P,P')})
U_0 (e^{2\pi i n_0/m},{\cal V}) 
\rangle
= 
e^{
-2\pi i\fr{n_0n_3}{k}\cdot \fr{k}{m}  \link ({\cal V}, ({\cal P,P'}))
}
\langle
 U_0 (e^{2\pi i n_0/m},{\cal V})
\rangle
$
implies 
the charge of $U_0$ under the 3-form symmetry is $-n_0 k/m$,
we find that 
the intersection of two $U_1$'s
becomes a symmetry generator of the 3-form symmetry
with the symmetry group $\bb{Z}_K$, 
where $K := k \cdot {p^2}/{\gcd(N,p^2)}$ with 
the nontrivial denominator of $N/p^2$.

The appearance of topological objects at intersections of symmetry generators is a signal of higher groups.
Our discussion in this Letter can have potential applications
to both of physics and mathematics.
For the physics side, we may apply this discussion to, e.g., topological superconductors.
To discuss 't Hooft anomalies would be an important issue to determine the ground state structures.
For the mathematics side, clarifying the precise definition of the 
4-group would lead to a construction of a new class of higher-groups.
We will address these issues in the forthcoming paper~\cite{HNY:inprep}.

\paragraph{Acknowledgements}
RY thanks Ryohei Kobayashi, Tatsuki Nakajima, Tadakatsu Sakai, and Yuya Tanizaki 
for helpful discussions. 
This work is supported in part by Japan Society of Promotion of Science (JSPS) Grant-in-Aid for 
Scientific Research (KAKENHI Grants No.~JP17H06462, JP18H01211 (YH), JP18H01217 (MN)
JP21J00480, JP21K13928 (RY)).

\appendix
\section{\label{charge}Charged objects for higher-form symmetries}
Here, we summarize charged objects for the higher-form symmetries.
The corresponding charged objects  are the Wilson loop 
and its analogues.
For 0-, 1-, 2-, and 3-form symmetries, 
the charged objects are explicitly written as
\begin{equation}
\begin{split}
&
L (q_0, {\cal P}) =  
e^{i q_0 \phi ({\cal P}) },
\\
&
W(q_1, {\cal C}) = e^{iq_1 \int_{\cal C} a} = 
U_2(e^{-2\pi i q_1/q},{\cal C}), 
\\
& V(q_2, {\cal S}) 
= e^{iq_2 \int_{\cal S} b}, 
\quad
 D(q_3, {\cal V}) = e^{i q_3 \int_{\cal V} c}, 
\end{split}
\end{equation}
respectively.
Here, the charges are quantized 
$q_0\in \bb{Z}_m$, $q_1\in \bb{Z}_p$, $q_2\in \bb{Z}_q$, and $q_3\in \bb{Z}_k$
because of the large gauge invariance for the gauge fields.
We remark that $W (q_1, {\cal C})$ is identical to the symmetry generator
$U_2(e^{-2\pi i q_1/q}, {\cal C})$.
The symmetry transformations are induced by the link of a charged operator and the symmetry generator (see \ref{corr}):
\begin{equation}
\begin{split}
& \vevs{  U_0 (e^{2\pi i n_0/m},{\cal V}) L (q_0, {\cal P}) }
= e^{2\pi i q_0 n_0 \link ({\cal V,P}) /m } \vevs{ L (q_0, {\cal P}) },
\\ 
& \vevs{  U_1 (e^{2\pi i n_1/p},{\cal S}) W (q_1, {\cal C}) }
= e^{2\pi i q_1 n_1 \link ({\cal S,C}) /p } 
\vevs{ W (q_1, {\cal C}) },
\\ 
& \vevs{  U_2 (e^{2\pi i n_2/q},{\cal C}) V (q_2, {\cal S}) }
= e^{2\pi i q_2 n_2 \link ({\cal C,S}) /q } 
\vevs{ V (q_2, {\cal S}) },
\\ 
& \vevs{  U_3 (e^{2\pi i n_3/q},({\cal P,P'})) D (q_3, {\cal V}) }
\\
&= e^{2\pi i q_3 n_3 \link ({\cal (P,P'),V}) /k } 
\vevs{D (q_3, {\cal V}) }.
\end{split}
\label{210516.1529}
\end{equation}

\section{Derivations of correlation functions\label{corr}}
Here, we briefly explain derivations of 
the correlation functions.
The derivations are based on the 
reparametrizations of dynamical fields in the path-integral formalism, which can be understood as 
finite versions of the Ward-Takahashi identity or 
Schwinger-Dyson equation.

As an example, we explain the derivation of
\er{210516.1648}.
In the path-integral formalism, 
the correlation function can be written as 
\begin{equation}
\begin{split}
&\vevs{  U_1 (e^{2\pi i n_1/p},{\cal S}) U_2  (e^{2\pi i n_2/q},{\cal C})}
\\
&=
{\cal N} 
\int {\cal D} [\phi, a, b, c] 
e^{iS_{\rm TAE}}
 U_1 (e^{2\pi i n_1/p},{\cal S}) U_2  (e^{2\pi i n_2/q},{\cal C}) ,
\end{split}
\end{equation}
where ${\cal N}$ is the normalization factor
such that $\vevs{1} = 1$.
We can absorb the symmetry generator $U_1$ 
into 
the action by the reparametrization of $a$
as follows.
We take a 3-dimensional subspace 
${\cal V_S}$ whose boundary is ${\cal S}=\partial {\cal V_S}$,
and express the integral in the symmetry generator
by the Stokes theorem
 as
 \begin{equation}
\int_{\cal S}
\left(\frac{q}{2\pi} b +\fr{N}{4\pi^2}\phi  da\right)
 =\int_{\cal V_S}
\left(\frac{q}{2\pi} db +\fr{N}{4\pi^2}d\phi \wed da\right).
 \end{equation}
By using this relation, 
one can absorb the symmetry generator 
to the action,
\begin{equation}
\begin{split}
&
e^{i S_{\rm TAE}[\phi, a,b,c] }
e^{-2\pi i \fr{ n_1}{p} \int_{\cal V_S}
 \left(\frac{q}{2\pi} db +\fr{N}{4\pi^2}d\phi \wed da\right)} 
\\
    &
    =
    e^{i S_{\rm TAE}\left[\phi, a +  \fr{2\pi n_1}{p}\delta_1 ({\cal V_S}),b,c\right]}.
\end{split}
\end{equation}
By the reparametrization 
$a +  ({2\pi n_1}/{p})\delta_1 ({\cal V_S})  \to a$
 in the path integral, 
 we obtain
 \begin{equation}
 \begin{split}
    &\vevs{
 U_1 (e^{2\pi i n_1/p},{\cal S}) U_2  (e^{2\pi i n_2/q},{\cal C})
}
\\
    &= 
e^{- \fr{2\pi i n_1 n_2}{p} \link ({\cal S,C}) }
  \vevs{U_2  (e^{2\pi i n_2/q},{\cal C} )}.
\end{split}
 \end{equation}
 Here,
 $\link ({\cal S,C})
  = \int_{\cal V_S} \delta_3 ({\cal C})$
  is the intersection number of ${\cal V_S}$
  and ${\cal C}$, which is equal to 
  the linking number of ${\cal S}$ and ${\cal C}$.

\end{document}